\documentclass[aps,prl,twocolumn,amsmath,amssymb]{revtex4-1}

\usepackage{amssymb}
\usepackage{amsmath}
\usepackage{graphicx}

\begin{document}

\title{Fast optical control of spin dynamics in a  quantum wire}

\author{Z.-G. Zhu, C.-L. Jia, and J. Berakdar}
\address{Institut f\"{u}r Physik, Martin-Luther-Universit\"{a}t
Halle-Wittenberg, 06099 Halle, Germany.}
% Nanotechnikum-Weinberg, Heinrich-Damerow-St. 4,
%06120 Halle, Germany;}

\begin{abstract}
The spin dynamics in a quantum wire with a Rashba spin orbit
interaction (SOI) is shown to be controllable via  sub-picosecond
electromagnetic pulses shaped appropriately. If the light polarization vector
is  along  the wire's direction, the carriers experience a momentum boost while
 the phase coherence in different spin channels is maintained, a fact  exploitable
to control the speed of  a photo-driven spin field effect transistor.
 A photon pulse with a polarization  vector
perpendicular to the  wire  results in a  spin
precession which is
comparable to that  due to the Rashba SOI and is tunable
by the pulse field parameters, an effect  utilizable in
optically controlled spintronics devices.
\end{abstract}

\pacs{85.75.Hh,73.63.Nm,85.30.Tv,73.63.-b} \maketitle

%\section{Introduction}
\section{Introduction}  Optical semiconductor devices are
indispensable components of nowadays technology  \cite{fukuda}
with applications  ranging from optical fiber communication
systems to consumer electronics. A new imputes is expected from spintronics devices
\cite{wolf,zutic}, i.e. from exploiting the carriers spin in addition to their charge
for efficient operation or to realize new
functionalities.
 While the  optical control and manipulation of
charges in conventional semiconductors \cite{neamen} have been the
key for the realization of ultra-fast electronic devices,
an analogous optical control of the spins is however
not straightforward.  The appropriate electromagnetic pulses
 are U(1) fields; while spins belong to SU(2) fields. A laser
pulse does not seem thus to couple to the spins directly, meaning
 a less efficient optical coupling to the spins than to the
charges. Indeed, the key to the optical spin-manipulation are
inherent interactions in the system that couple the charge to the
spin such as SOI \cite{spin,zhu08}.
 Along this line we
 present here a way for an ultrafast control of the spin dynamics in
a conventional spin field effect transistor  (SFET) \cite{datta,koo} driven
by shaped electromagnetic pulses.
The  SFET relies on  the Rashba SOI to perform a
controlled rotation of a carrier spin that  traverses
a FET-type device \cite{datta,koo} with two magnetic leads (cf. Fig.\ref{fig11}).
The conductance depends on the achieved rotation angle $\Delta\theta_0$ at the drain lead.
Here we propose two ways to control the time-dependence of this rotation angle
and hence the operation of SFET.
The key ingredient is the use of
asymmetrically shaped  linearly  polarized
electromagnetic pulses
\cite{tielking,hcp1,hcp2,matos}. The pulse consists of a very short, strong half-cycle
followed by a second long (compared to the ballistic transverse time)
and a much weaker half-cycle of an
opposite polarity \cite{tielking,hcp1,hcp2}. Hence such pulses
are often called half-cycle pulse (HCP).
 Experimentally the achieved asymmetry ratio of the
positive and negative amplitudes can be 13:1, the peak fields can
reach several hundreds of kV/cm and have a duration
$t_{\text{p}}$ in the range
between nano and subpicoseconds.
The interaction of HCP with matter is particular in that it delivers
a definite amount of momentum boost to the carriers along the optical
polarization axis \cite{matos,zhu08}. In case this axis is along the conductive channel
we find that both the spin precession frequency and the
carrier speed  increase upon irradiation but  $\Delta\theta_0$ remains unchanged.
The operation speed is thus pulse-controllable.
 If the optical polarization vector is perpendicular to the carriers propagation direction
 $\Delta\theta_0$ and hence the SFET operation is determined by the pulses parameters,
 as shown explicitly below.\\
\section{Model SFET}
\begin{figure}[tbph]
\vspace*{33pt}
%\centering \includegraphics[width=4 cm, height=3.2 cm,angle=-90]{fig1}
\centering \includegraphics[width=0.15\textwidth,angle=-90]{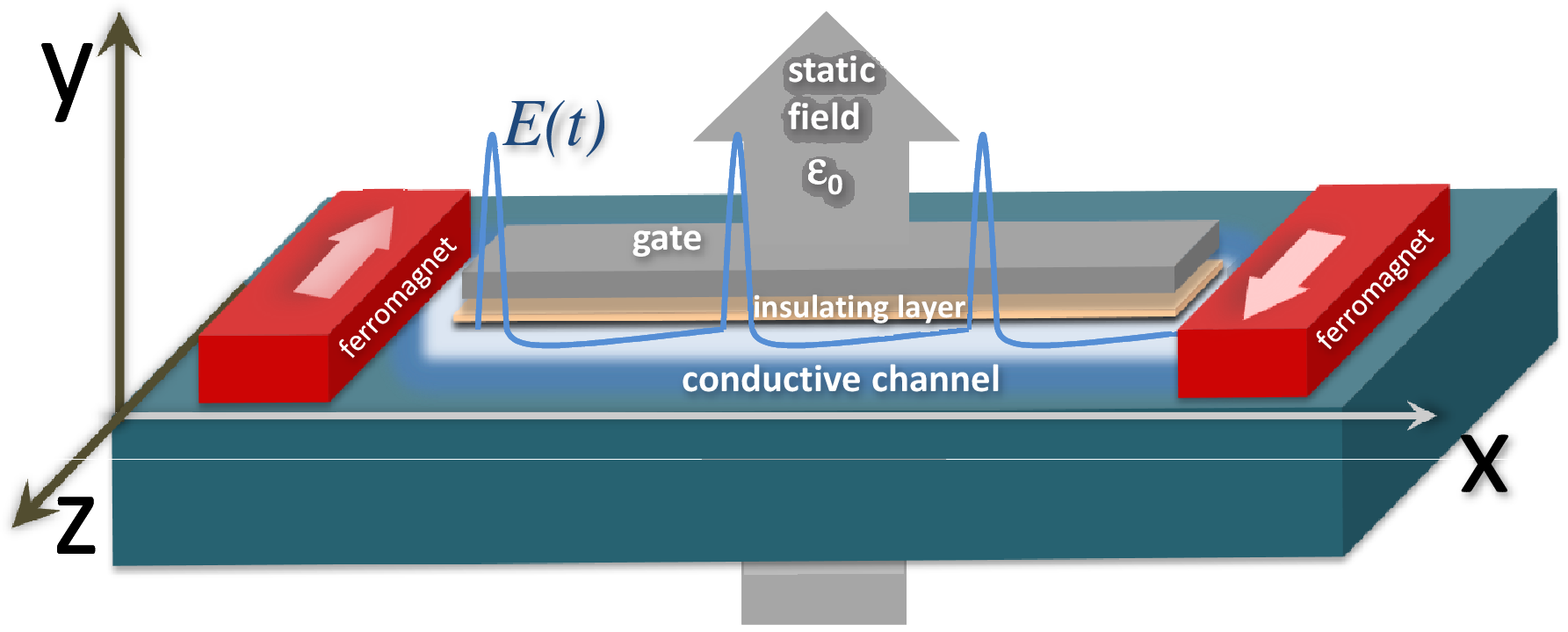}
\caption{Schematics of the optically driven spin field transistor.
 Ferromagnetic leads are separated from the conductive channel
by a tunneling barrier to enhance the spin injection efficiency. A
 metallic gate is used to tune the Rashba SOI via a static field $\epsilon_0$. $E(t)$ is the time-dependent
 electric field. }
\label{fig11}
\end{figure}

We follow Refs.[\onlinecite{datta,koo}] and
focus on the central conductive region of the SFET that can be considered as a one dimensional (1D)  quantum wire
(QW) \textit{of length $L$} with  a spin orbit interaction (SOI).  The ferromagnetic
leads serve as spin injector and spin detector separated from QW
by an insulating barrier to achieve a higher spin injection efficiency (Fig.\ref{fig11}).
Recently a  SFET structure similar to Fig.\ref{fig11} has been   realized experimentally \cite{koo}.
%The results compare favorably with the theory of \cite{datta,koo}, i.e. with the stationary version of the present
%work.
The experimental findings are in line with the predictions of  the stationary version of the present work.
In the work presented below, however, we consider 1D QW, for  spin-flip transitions between the first and second subband is negligible
for the system studied in this work.
%
%%%%%%%%%%%%%%%%%%%%%%%%%%%%%%%%%%%%%%%%%%%%%%%%%%%%%%%%%%
%%%%%%%%%%%%%%%%%%%%%%%%%%%%%%%%%%%%%%%%%%%%%%%%%%%%%%%%
%

The inversion-asymmetry of the confining potential results in
 Rashba SOI \cite{rashba}
 $H_{\text{R}}$.  For a two-dimensional electron gas  $H_{\text{R}}$ reads \cite{rashba}
\begin{equation}
H_{\text{R}}=\alpha^{0}_{\text{R}}(\sigma_{z}k_{x}-\sigma_{x}k_{z}),
\label{rashba1}
\end{equation}
where $\sigma_{i}, i=x,y,z$ are Pauli matrixes,
$\alpha^{0}_{\text{R}}=r_{\text{R}}\varepsilon_{0}$ is the static Rashba SOI coefficient
which is proportional to the perpendicular electric field
$\varepsilon_{0}$ resulting from
band bending, $r_{\text{R}}$ is a material-specific
prefactor \cite{winkler}.
Under equilibrium conditions, spin transport in such a
device is investigated extensively \cite{datta,winkler}; in brief, one chooses for  1D
QW the $z$ axis as the spin quantization axis (Fig\ref{fig11}), meaning that
%\begin{equation}
$$H_{\text{R}}=\alpha^{0}_{\text{R}}\sigma_{z}k_{x}=\mu_{\text{B}}\sigma_{z}\tilde{B}_{z},
$$%\label{hr1}
%\end{equation}
where $\mu_{\text{B}}$ is Bohr's magneton, $\tilde{B}_{z}=\alpha^{0}_{\text{R}}k_{x}/\mu_{\text{B}}$ is an effective
magnetic field along $z$. This results
in the spin splitting $2\alpha^{0}_{\text{R}}k_{x}$ between carriers injected
with spin polarization parallel or antiparallel to $z$.
The phase difference while passing through the
 length $L$ is $$\vartriangle\theta_{0}=(k_{x\uparrow}-k_{x\downarrow})L=\frac{2m^{*}\alpha^{0}_{\text{R}}L}{\hbar^{2}}.$$ The eigenenergies and the eigenstates are respectively $$E_{k\mu}=\frac{\hbar^{2}k^{2}}{2m}+\mu\alpha_{\text{R}}^{0}k$$ and  $$|k\mu\rangle=\frac{1}{\sqrt{L}}e^{ikx}|\eta_{\mu}\rangle$$ where $|\eta_{\mu}\rangle$ is the spin states.
This is the original idea of the Ref. \cite{datta}.\\
Considering the spins to  be injected  aligned along the  $x$ or $y$ directions, they precess around
$\tilde{B}_{z}$. In Heisenberg picture, the spin operators vary with time as
%\begin{equation}
$$\dot{\sigma}_{x(y)}(t)=\mp\omega_{k_{x}}\sigma_{y(x)}(t), $$
% \notag \\
%\frac{d\sigma_{y}(t)}{dt} &=&\omega_{k_{x}}\sigma_{x}(t),
%\label{sigmaxy}
%\end{equation}
where  $$\omega_{k_{x}}=\frac{2\alpha^{0}_{\text{R}}k_{x}}{\hbar}$$ is the precession frequency. Therefore,
 $$\sigma_{\pm}(t)=\sigma_{x}\pm i\sigma_{y}=\sigma_{\pm}(0)e^{\pm i\omega_{k_{x}}t}.$$
Let us specify the initial orientation, say
$\sigma_{y}(0)=0$, therefore $$\sigma_{x}(t)=\sigma_{x}(0)\cos(\omega_{k_{x}}t) \mbox{ and }
\sigma_{y}(t)=\sigma_{x}(0)\sin(\omega_{k_{x}}t).$$  % which is schematically shown in Fig. 2(a)
An initial spin along the $x$ direction  rotates anticlockwise with the angular frequency
$\omega_{k_{x}}$. The accumulated angle through the length  $L$ is
%\begin{equation}
$\vartriangle\theta_{0}=\frac{2m^{*}\alpha^{0}_{\text{R}}L}{\hbar^{2}}$,
%\label{theta0}
%\end{equation}
which is exactly equal to the phase shift for the spin along $z$.
% The meaning of $\vartriangle\theta_{0}$
% in $x$ and $y$ polarized spin cases is the accumulated procession angle.

%%%%%%%%%%%%%%%%%%%%%%%%%%%%%%%%%%%%%%%%%%%%%%%%%%%%%%%%%%%%%%%%%%%%%%%%%%%%%%%%%%%%%%%%%%%%%%%%%
%%%%%% The First Case %%%%%%%%%%%%%%%%%%%%%%%%%%%%%%%%%%%%%%%%%%%%%%%%%%%%%%%%%%%%%%%%%%%%%%%%%%%%
%%%%%%%%%%%%%%%%%%%%%%%%%%%%%%%%%%%%%%%%%%%%%%%%%%%%%%%%%%%%%%%%%%%%%%%%%%%%%%%%%%%%%%%%%%%%%%%%%%

%\section{The model with pulses}
\section{The first dynamic case}
Having outlined the equilibrium case, we apply to the quantum wire  a
 linearly polarized HCP  with the vector potential
$\mathbf{A}=\mathbf{e}_{x}A(t)$.
The polarization
 vector  $\mathbf{e}_{x}$  is along the $x$ direction. Thus,
%$\mathbf{B}=0$,
$$\mathbf{E}(t)=-\mathbf{e}_{x}\frac{\partial A(t)}{\partial t}=Fa(t)\mathbf{e}_{x}$$
%where $\mathbf{e}_{x}$ is the unit vector of
%$\mathbf{E}(t)$,
where $F$ is the peak amplitude of the electric field and
$a(t)$ describes the  pulse profile.
The single particle Hamiltonian reads
\begin{equation}
H=\frac{\boldsymbol\pi^{2}}{2m^{*}}+V(y,z)-e\Phi+
\frac{\alpha^{0}_{\text{R}}}{\hbar}(\boldsymbol\sigma\times\boldsymbol\pi)_{y,k_{z}=0},
\label{Hd1}
\end{equation}
where $\boldsymbol\pi=\mathbf{p}+e\mathbf{A}(\mathbf{r},t)$, and
$\mathbf{p}=-i\hbar\triangledown$ is the momentum operator.
%$\mathbf{A}$ is the vector potential of the laser pulse at
%$\mathbf{r}$, at time $t$.
The second term in Eq. (\ref{Hd1}) is
the QW confinement potential, the third term is from the scalar
potential of the pulse field, and the fourth term in Eq. (\ref{Hd1})
is the Rashba SOI. % which presents in the 1D quantum wire.
We write
$H=H^{0}+H^{t}$,
with
%\begin{eqnarray}
$$H^{0}=\mathbf{p}^{2}/2m^{*}+V(y,z)+(\alpha^{0}_{\text{R}}/\hbar)(\boldsymbol\sigma\times\mathbf{p})_{y,\text{
} k_{z}=0},$$ and
%\notag\\
\begin{eqnarray}
H^{t}&=&(e/2m^{*})(\mathbf{p}\cdot\mathbf{A}+\mathbf{A}\cdot\mathbf{p})+(e/2m^{*})\mathbf{A}\cdot\mathbf{A}-
e\Phi(\mathbf{r}) \notag \\
&+&(e\alpha^{0}_{\text{R}}/\hbar)(\boldsymbol\sigma\times\mathbf{A})_{y}.
\notag
\end{eqnarray}
%\label{hi0t}
%\end{eqnarray}
 $H^{0}$ is the pulse-free single particle Hamiltonian, and $H^{t}$ is the time-dependent
part. Choosing a gauge where $\Phi=0$;
$\mathbf{p}\cdot\mathbf{A}+\mathbf{A}\cdot\mathbf{p}=2A_{x}p_{x}$
and $(\boldsymbol\sigma\times\mathbf{A})_{y}=\sigma_{z}A_{x}$ results
in
\begin{equation}
H^{t}=\frac{e}{m^{*}}A_{x}p_{x}+\frac{e^{2}}{2m^{*}}A^{2}_{x}+\frac{e\alpha^{0}_{\text{R}}}{\hbar}\sigma_{z}A_{x}.
\label{hit}
\end{equation}
Defining the spinor field operator as
%\begin{equation}
$\hat{\psi}(x)=\sum_{k\mu}c_{k\mu}|k\mu\rangle$,
%\label{fieldop}
%\end{equation}
where $c_{k\mu}$ is the annihilation operator for the states $|k\mu\rangle$,
in the second quantization form, we obtain
\begin{equation}
H^{t}=\sum_{k\mu}\left(\frac{e\hbar}{m^{*}}A_{x}k+\frac{e^{2}}{2m^{*}}A^{2}_{x}+
\mu\frac{e\alpha^{0}_{\text{R}}}{\hbar}A_{x}\right)c^{\dagger}_{k\mu}c_{k\mu}.
\label{hit1}
\end{equation}
The total Hamiltonian reads
%\begin{equation}
$H=\sum_{k\mu}\varepsilon^{c}_{k\mu}(t)c^{\dagger}_{k\mu}c_{k\mu}$,
%\label{hc}
%\end{equation}
with the time-dependent transient energy (measured with respect to the ground  state of $V(y,z)$)
\begin{equation}
\varepsilon^{c}_{k\mu}(t)=\frac{1}{2m^{*}}[(\hbar
k+eA_{x}(t)+\mu\frac{p_{\text{so}}}{2})^{2}-\frac{p^{2}_{\text{so}}}{4}],
\label{cenergy}
\end{equation}
and
\begin{equation}
p_{\text{so}}=\hbar k_{\text{so}}=\frac{2m^{*}\alpha^{0}_{\text{R}}}{\hbar}.
\label{pso}
\end{equation}

To obtain  the momentum distribution upon a short pulse
application (say at $t=0$) we may proceed as in \cite{zhu08,matos} and
expand the single-particle excited state
$\Psi_{k_{0}\mu_{0}}(x,t)$ that starts from the state labeled by $|k_{0}\mu_{0}\rangle$ in terms of the stationary eigenstates
\begin{equation}
\Psi_{k_{0}\mu_{0}}(x,t)=\sum_{\mu}\int dk C_{k\mu}(k_{0},\mu_{0},t)|k\mu\rangle e^{-iE_{k\mu}t/\hbar}.
\label{wfa}
\end{equation}
For a sudden-excitation, $\Psi_{k_{0}\mu_{0}}(x,t=0^{+})$ right after the pulse evolves from the state before the pulse
$\Psi_{k_{0}\mu_{0}}(x,t=0^{-})$ as \cite{zhu08}
$$\Psi_{k_{0}\mu_{0}}(x,t=0^{+})=e^{ix\bar{p}} \Psi_{k_{0}\mu_{0}}(x,t=0^{-}).$$ Thus
%for a sudden-excitation , we get %i.e. if the duration $t_{p}$ of the positive-polarity half cycle of the pulse
 %is less than the ballistic time (a situation realizable experimentally),
\begin{equation}
C_{k\mu}(k_{0},\mu_{0},t=0^{+})=C_{k-\bar{k},\mu}(k_{0},\mu_{0},t=0^{-}),
\label{cparameter}
\end{equation}
where $$\hbar\bar{k}=\bar{p}=-eF\int_{-t_{p}/2}^{t_{p}/2} \;a(t') \,{d}t'$$ is the momentum boost delivered to the carrier by the pulse (the second weak and long half cycle of the pulse acts as a weak DC off-set field).
For $t<0$,  eq.(\ref{wfa}) may stand for  the injected electron state in terms of the stationary states. For  $C_{k'\mu'}(k,\mu,t<0)=\delta(k-k')\delta_{\mu\mu'}$,  the injected electron  occupies a single eigenstate. In this case
the wave function  after the pulse is
$$\Psi_{k_{0}\mu_{0}}(x,t>0)=e^{i(\hbar k_{0}+\bar{p})x}|\eta_{\mu_{0}}\rangle e^{-iE_{(k_{0}+\bar{k})\mu_{0}}t/\hbar}/\sqrt{L}.$$
The energy after the pulse is
$$\varepsilon^{c}_{k\mu}(t>0)=\frac{1}{2m^{*}}\left[(\hbar
k+\bar{p}+\mu\frac{p_{\text{so}}}{2})^{2}-\frac{p^{2}_{\text{so}}}{4}\right].$$
%
%
%follows from (\ref{cenergy}) by replacing $eA_{x}$ by $\bar{p}$.
If $C_{k'\mu'}(k,\mu,t<0)$ models a Gaussian wave packet centered at $k_{0}$ for the injected state we infer from  Eq. (\ref{cparameter}) that right after the pulse the shape of the wave packet is maintained  while its central momentum is shifted by
 $\bar{p}$.
We conclude:\\
 i) The pulse field
delivers a transient momentum transfer (which is proportional to the momentary vector potential) and a net
momentum given by the field-amplitude time-integrated  over the field duration, for harmonic
fields this quantity vanishes whereas for HCP it is finite and is equal to $\bar{p}$.\\
ii) This momentum boost is experienced by all the electrons  speeding up
 the device operation.\\
  iii) The phase difference $\Delta \theta_0$ is
maintained as for the static case, i.e.\ the operation speed is changed by an amount
 proportional  to the field strength while
 the spin coherence is unchanged, a fact exploitable for the realization of an
ultra-fast SFET.

%%%%%%%%%%%%%%%%%%%%%%%%%%%%%%%%%%%%%%%%%%%%%%%%%%%%%%%%%%%%%%%%%%%%%%%%%%%%%%%%%%%%%%%%%%%%%%%%%%%%%%%%%%%%%
%%%%%%%%%%%%%%%%%%%%%%%%%%%%%%%%%%%%%%%%%%%%%%%%%%%%%%%%%%%%%%%%%%%%%%%%%%%%%%%%%%%%%%%%%%%%%%%%%%%%%%%%%%%%%

\section{The second dynamic case}
If the electric field polarization of the HCP is along the
$y$ direction, i.e.   perpendicular to the 2DEG the Hamiltonian is
\begin{equation}
H=\frac{\boldsymbol\pi^{2}}{2m^{*}}+V(\mathbf{r})-e\Phi+
\frac{\tilde{\alpha}(t)}{\hbar}(\boldsymbol\sigma\times\boldsymbol\pi)_{y,k_{z}=0}, % \notag \\
% &\tilde{\alpha}(t)&=\alpha^{0}_{\text{R}}+\alpha^{t}_{\text{R}}(t),
\label{Hd2}
\end{equation}
where $$\tilde{\alpha}(t)=\alpha^{0}_{\text{R}}+\alpha^{t}_{\text{R}}(t),$$ and $\alpha^{0}_{\text{R}}$ is the static Rashba SOI.
$$\alpha^{t}_{\text{R}}(t)=r_{\text{R}}E(t)$$ is proportional to the
HCP electric field. The time
dependent part of the Hamiltonian is
\begin{equation}
H^{t}=\frac{e^{2}}{2m^{*}}A^{2}_{y}+\frac{\alpha^{t}_{\text{R}}(t)}{\hbar}(\boldsymbol\sigma\times\mathbf{p})_{y}.
\label{ht2}
\end{equation}
The vector potential in the canonical momentum in the second
term in Eq. (\ref{ht2}) does not couple to the electric field of
the HCP. The total Hamiltonian is $H=H^{0}+H^{t}$.  The first term in Eq. (\ref{ht2})
results in a phase shift for all states in all subbands, an effect which
is unimportant for the
following discussion and hence we ignore it and consider  the total Hamiltonian
\begin{equation}
H=\frac{\hbar^{2}k^{2}_{x}}{2m^{*}}+\frac{\tilde{\alpha}(t)}{\hbar}(\boldsymbol\sigma\times\mathbf{p})_{y}.
\label{h2}
\end{equation}
%%%%%%%%%%%%%%%%%%%%%%%%%%%%%%%%%%%%%%%%%%%%%%%%%%%%%%%%%%
\begin{figure}[tbph]
%\centering \includegraphics[width =6.5 cm, height=5.2 cm]{fig2}
\centering \includegraphics[width=0.38\textwidth]{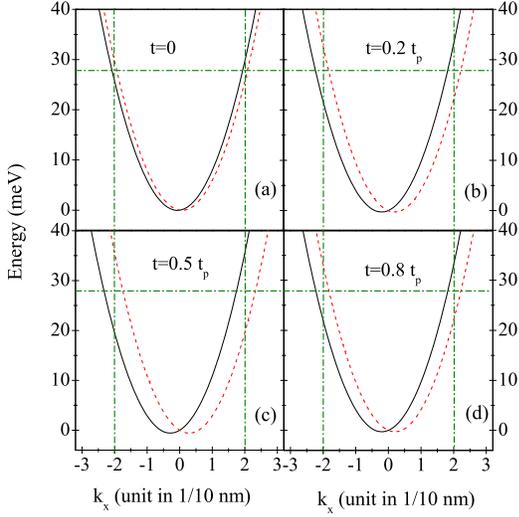}
\caption{The instantaneous energy spectrum for GaSb/InAs/GaSb system \cite{luo}  at different
 time fraction of the pulse duration $t_p$. In (a) the static spectrum with SOI splitting is shown. The parameters are taken from Ref. \cite{luo}: $m^{*}=0.055 m_{0}$, $\alpha^{0}_{R}=0.9\times 10^{-9} eV.cm$, $2\alpha^{0}_{R}k_{f}=4.0 meV$, $k_{f}=2.0 (10nm)^{-1}$. In (b-d) the horizontal dash-dot lines indicate the position of the Fermi level for equilibrium system, and the vertical dash-dot lines mark the positions of the Fermi wave vectors. The solid (dash) curves correspond to the spin up (down) state. The other parameters are selected as $\frac{F}{\varepsilon_{0}}=3$, $a(t)=sin(\pi t/t_{p})$ for $0\leq t\leq t_{p}$. } \label{fig2}
\end{figure}
%%%%%%%%%%%%%%%%%%%%%%%%%%%%%%%%%%%%%%%%%%%%%%%%%%%%%%%%
%We get instantaneous energy from the above equation.
We note, the pulse field does not change the quantum number $k$ nor the spin states.
The time evolution  occurs only in the parameter space specified by  $\tilde{\alpha}(t)$ which is 1D parameter space. With varying $\alpha^{t}_{\text{R}}(t)$ the spin-dependent potential is changed and so does the energy.
To be specific let us consider the  GaSb/InAs/GaSb system with the parameters given in Ref. \cite{luo}.
For InAs QW with a width of 6.28 nm
the second energy subband is separated from the ground state  by $\approx$ 700 meV  \cite{luo}.
To inspect the effect of the pulse field we show
in Fig. \ref{fig2} the instantaneous energy spectrum.
As evident from this figure the highest achieved energy level is well below the first excited subband
and hence we only need to consider the intra-band dynamics in the first subband.
 The behavior of the instantaneous energy for a state $|k\mu\tilde{\alpha}(t)\rangle$ is such
 that  for positive $k$ the energy of the spin up carrier  is raised while the spin-down energy is lowered in the first quarter of the mono-cycle pulse; the opposite happens  in the second quarter cycle. In the second half of the mono cycle  the spin-resolved energies evolves in an opposite way to the first half.
 As the shift is determined by the magnitude of the   electric field peak, the very weak and long (on the transport time) tail of HCP has a minor effect. From Fig. \ref{fig2} we infer that  the pulse results in a time-varying potential and
  hence an oscillation of all electrons in the Fermi sphere in energy space.
  No holes are generated.
The spin operators develop as
%\begin{equation}
$ \dot{\sigma}_{\pm}(t)=\mp i\tilde\omega_{k_{x}}(t)\sigma_{\pm}(t)$,
%\label{sigmapm}
%\end{equation}
where $\tilde\omega_{k_{x}}(t)=2\tilde\alpha(t)k_{x}/\hbar$. Hence, we find
\begin{equation}
\sigma_{\pm}(t)=\sigma_{\pm}(0)\exp\Bigg{\{}\pm i\int\tilde\omega_{k_{x}}(t)dt\Bigg{\}}.
\label{sigmapmt}
\end{equation}
Defining
%\begin{equation}
$\tau=-(F/\varepsilon_{0})t_{\text{p}}\gamma$,
%\label{tau1}
%\end{equation}
where $\gamma=\int a(\xi')d\xi'$, and $\xi'=t'/t_{\text{p}}$.
%If we define $\tau$ as
%\begin{equation}
%\tau=-\frac{F}{\varepsilon_{0}}\int^{\infty}_{-\infty}a(t')dt',
%\label{tau}
%\end{equation}
 we obtain
%\begin{equation}
$\int\tilde\omega_{k_{x}}(t)dt=\omega_{k_{x}}t+\theta_{\text{p}}$,
%\label{intomega}
%\end{equation}
where
\begin{equation}
\theta_{\text{p}}=-\omega_{k_{x}}\tau
\label{thetap}
\end{equation}
%The integral in $\gamma$ is taken really over the time
%range of the HCP pulses, however, the extension to infinity
%doesn't affect the result.
For an injected carrier with a spin polarization vector along $z$, the wave function
right after the pulse $\Psi_{k_{\mu_{0}}\mu_{0}}(x,t>0)$ evolves
from the stationary state before the pulse $\Phi_{k_{\mu_{0}}\mu_{0}}(x,t<0)$ as
% can obtain the wave function match condition based on the time-evolution operator
% theory \cite{matos} which is similar to Eq. (\ref{match}) but the factor $e^{ix\bar{p}}$ is
% replaced by $e^{-is_{0}\theta_{p}/2}$.
%
$$\Psi_{k_{\mu_{0}}\mu_{0}}(x,t>0)=e^{-i\mu_{0}\theta_{p}/2}\Phi_{k_{\mu_{0}}\mu_{0}}(x,t<0),$$
meaning that
the pulse causes a  phase splitting  in the spin up and down channels. The phase
difference between up and down spins is  equal to the induced angle rotation in Eq. (\ref{thetap}).\\
\section{Optical SFET}
\begin{figure}[tbph]
\centering \includegraphics[width =7.5 cm, height=3.2 cm]{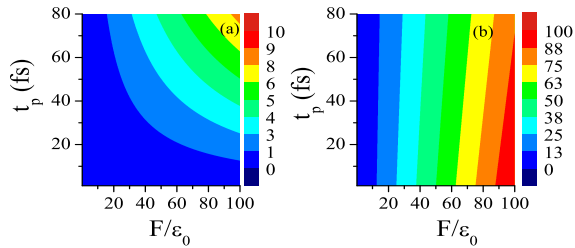}
%\centering \includegraphics[width=0.75\textwidth]{fig3}
\caption{A contour plot for the ratio $\lambda$, as given by  eq.(\ref{lamda}),  as a function of
 the magnitude and of the duration of the external field for a single HCP  (a) and  for a train of HCPs  (b).
 In (a), the Fermi velocity is chosen as 0.4 $\mu$m/ps, L=0.2 $\mu$m, $\gamma$ is set at 0.5. In (b), $t_{\text{s}}=0.5$ ps, $t_{\text{tr}}=0.5$ ps, $\gamma\approx1$.} \label{fig3}
\end{figure}

In the static case a  0.67 $\mu m$ long 1D quantum wire is needed to reach the phase shift $\vartriangle\theta_{0}=\pi$ in 2D In$_{x}$Ga$_{1-x}$As \cite{das}. While a shorter 1D quantum wire is sufficient to reach $\pi$ phase shift with length 0.2 $\mu m$ in
GaSb/InAs/GaSb system \cite{luo}, since the Rashba SOI is larger, i.e. $\alpha^{0}_{R}\approx 0.9\times 10^{-9} eV cm$. According to Ref. \cite{luo} the charge density is $n=10^{12} cm^{-2}$, and $2\alpha^{0}_{R}k_{f}=4\, meV$. We find then  $\omega_{k_{x}}\approx 2\pi\, ps^{-1}$.  The Fermi velocity is  $0.4 \mu m/ps$.
Thus in 0.2 $\mu m$ 1D QW there are 20 electrons distributed and the transport time $t_{\text{tr}}$
for the electrons at the Fermi velocity is about 500 fs. In what follows we based our discussion on these
 realistic numbers.
HCPs with peak field of up to several hundreds of kV/cm and duration in the picosecond and subpicosecond regimes can be experimentally generated \cite{hcp1}. Novel principles allow the generation of unipolar pulses as short as 0.1 fs and with intensities up to 10$^{16}$ W/cm$^{2}$  \cite{hcp2}.
The pulse induced precession
angle is
%\begin{equation}
%\theta_{\text{p}}&=&\frac{2m^{*}\alpha^{0}_{\text{R}}L_{\text{p}}}{\hbar^{2}}\Big{(}\frac{F\gamma}{\varepsilon_{0}}\Big{)}, \notag \\
$$\theta_{\text{p}}=k_{\text{so}}L_{\text{p}}(F\gamma/\varepsilon_{0}),$$
%\label{thetap1}
%\end{equation}
where $L_{\text{p}}$ is the  length traveled by a particle with a momentum $\hbar k_{x}$  within  the pulse duration $t_{\text{p}}$.
The total precession angle accumulated while traversing the length $L$
is $$\vartriangle\theta=\vartriangle\theta_{0}+\theta_{\text{p}},$$ where the
first term stems from the static Rashba SOI. To evaluate
the angle $\theta_{\text{p}}$ we consider the ratio
\begin{equation}
\lambda=\frac{\theta_{\text{p}}}{\vartriangle\theta_{0}}=\frac{L_{\text{p}}}{L}\frac{F\gamma}{\varepsilon_{0}}.
\label{lamda}
\end{equation}

i) Single HCP.  For $t_{p}=20 fs$ we note  $t_{p}\ll t_{\text{tr}}$, and
 % $a(\xi')$ in this case can be regarded as a delta function (if it is assumed to be a Gaussian function, the result % is not changed),
 we have $\gamma\approx 1$.
The static fields $\varepsilon_{0}$ are typically
of the order of several $kV/cm$, for example, in
GaAs/Al$_{0.3}$Ga$_{0.7}$As quantum well
\cite{winkler}. $F$ can be
generated with several hundreds of $kV/cm$. Therefore,
$F/\varepsilon_{0}$ can be tuned as high as 100 (without inducing  inter-subband transitions).
$$L_{\text{p}}=\hbar k_{x}t_{\text{p}}/m^{*}$$ is about 8 nm during $t_{p}=20 fs$. Therefore, $\lambda\approx$  4.
The induced accumulated angle $\theta_{\text{p}}$ is a swift procession angle
transfer. If the initial injected spin is in $x$ direction it
suddenly rotates anticlockwise over an angle $\theta_{\text{p}}$ upon
applying the pulse. This conclusion is exploitable to realize
a nanosize SFET.
Spins with higher drift velocity experience a larger angular transfer during the period $t_{\text{p}}$. Thus,
a wave packet of spins   that is initially polarized in the same direction but contain
 different velocity components will
 spread over a range of angles after
scattering from the pulses. This can be compensated by operating the device  in the linear response
regime (small bias) where  electrons at the Fermi surface  dominate the
transport, i.e. $k_{x}=k_{f}$. A contour plot of the ratio $\lambda$, as introduced in
 eq.(\ref{lamda}), as function of the external field parameters is shown in Fig. \ref{fig3}(a).  The ratio increases with increasing $F$ and $t_{p}$. %The dependencies of $\lambda$ on these two parameters are inverse with respect to each other.}

%%%%%%%%%%%%%%%%%%%%%%%%%%%%%%%%%%%%%%%%%%%%%%%%%%%%%%%%%%
%%%%%%%%%%%%%%%%%%%%%%%%%%%%%%%%%%%%%%%%%%%%%%%%%%%%%%%%

ii) A train of HCPs: We apply a pulse train with a total duration  $t_{\text{t}}=n(t_{p}+t_{\text{s}})$, where $n$ is the number of HCP peaks, $t_{\text{s}}$ is the time interval between two consecutive  HCPs. $m=t_{\text{t}}/t_{\text{tr}}$ is the number of electrons passing through the device during $t_{\text{t}}$. We define for a single electron
an  averaged $\bar{\gamma}=n\gamma/m$   and  $L_{\text{p}}=L$.
 For  $t_{p}=1$ ps, $t_{\text{s}}=0.5$ ps, $t_{\text{t}}=125$ ps, $n=50$, $m=250$, $\gamma\approx1$, thus $\bar{\gamma}\approx0.2$ and $\lambda\approx20$. In
this case, spins with higher velocities will pass through the
device faster while  rotating faster.
%The extra precession
%angles are all the same which is determined by the length $L$ and independent of $k_{x}$ which
%is the similar to the static case in Eq. (\ref{theta0}).
The additional precession angle induced by the pulse field is comparable to
the static one or may  even be larger. To show the behavior more clearly, a contour plot of $\lambda$ with the tuning parameters of the external field is given in Fig. \ref{fig3}(b). Since the transport time is fixed, increasing $t_{p}$ decreases the ratio  $\lambda$ which is in a contrast to the single HCP case. However, a larger $\lambda$  is also acheived
via increasing the electric field strength.
%Another observation is that the single duration is not suffering from any limitation and the ratio $\lambda$ is enhanced by the train of HCPs in this case. }
As discussed for the static case, an injected spin polarized in
positive or negative $z$ direction  experiences a phase shift, while passing through the
length $L$,  that is exactly equal to the precession angles for the case of a spin injected
polarized along $x$ or  $y$.

%ii) The coupling between
%the vector potential and the spin provides an effective magnetic
%field (EMF) in $z$ direction which is proportional to $A_{x}$ (which
%is actually the transferred momentum), leading to enhanced
%precession of spins. This EMF can be controlled by the laser pulse
%and can be larger than the EMF generated by the SOI in principle.
%iii)If $\mathbf{A}$ only has the component in $y$ direction
%(perpendicular to the plane of 2DEG), then
%$\mathbf{P}\cdot\mathbf{A}+\mathbf{A}\cdot\mathbf{P}=0$,
%$(\vec{\sigma}\times\mathbf{A})_{y}=0$, and
%$h^{t}_{i}=\frac{e^{2}}{2m^{*}}A^{2}_{y}$. Thus this pulse only
%leads to an energy shift. Nothing special happens.
%iv)

%If $\mathbf{A}$ only has the component in $z$ direction,
%$\mathbf{P}\cdot\mathbf{A}+\mathbf{A}\cdot\mathbf{P}=0$,
%$(\vec{\sigma}\times\mathbf{A})_{y}=-\sigma_{x}A_{z}$, and
%$H^{t}=\frac{e^{2}}{2m^{*}}A^{2}_{z}+\frac{e\alpha_{R}}{\hbar}\sigma_{x}A_{z}$.
%In this case, the effective magnetic field is in $x$ direction, which is different to that
%caused from the Rashba SOI that is in $z$ direction. So the total effective magnetic field
%is laying in $xz$ plane and violates from $x$ and $z$ directions. The spin
%precession should be curved in the plane and there is no good
%quantum number to character the motion of spins.

\section{Experimental realization}
To be specific we propose a  GaSb/InAs/GaSb  structure (as realized in Ref. \cite{luo})
with InAs QW  width of 6.28 nm and a length of $\approx 100$ nm driven by pulses with a stength of 10 kV/cm and duration of 1 ps.
The nonzero $k_{z}=\pi/w$ leads to a spin-flip transition between subbands with a magnitude linear in $k_{z}$; while the energies for subbands are proportional to $k^{2}_{z}$.  In this configuration our model is  valid since $w/L<<1$.
%Thus the ratio $\Delta\theta w/3\pi L$, for , justifies the transitions negligible.}
For smaller $L$, we reduce $w$. The spin-precession is however controlled by the pulses and hence our SFET is still operational.
%Weak asymmetric
%pulses allow for the realization of considerably smaller optical SFET with
%a length well below the . Hence,
 For $L$  is less than the mean-free-path, dephasing
caused by impurities is negligible. Scattering from (acoustic) phonons
is spin independent causing a current relaxation
within tens of picoseconds \cite{moskalenko} at few Kelvins which is
larger than our transport time.
% Additionally, the carrier-phonon
%interaction is spin independent. Thus, the  spin coherence  is maintained and,
%as discussed above, the rotation angle $\Delta\theta_0$ is not affected even though
%the device becomes slower.
 Since the pulses are weak
the second transverse subband is not reached, hence
inter-subband transitions play no role.  Also, multiphoton processes are subsidiary for such weak pulses.
Finally we remark, symmetric pulses do not
lead to a net currents ($\gamma\equiv 0$) whereas a similar effect is achievable via
 quantum interferences in (the higher-energy) inter-band, one-photon, two-photon
 absorption \cite{corkum}. Our photo-induced current for the first dynamic is sizable:
  For the above device based on GaSb/InAs/GaSb the static current is usually $\approx 1.07 \; \mu$A. The ratio between the induced  and the static current is $\Delta=2\bar{p}/\hbar k_{f}$ which  varies in a large range. For F=10 kV/cm and $t_{p}=1$ ps we find $\Delta=7.6$ which means an induced current of $\approx 8 \, \mu$A.
%
%In this letter, we investigate such a phenomena where spin dynamics in so-called
%spin field transistor can be controlled directly by laser pulse which may open
%a way to the optically controlled spin dynamics in the future applications. We consider
%one kind of special laser pulse which is named as strongly asymmetric monocycle
%linearly polarized electromagnetic pulse (i.e. HCP). A peculiar property of HCP
%is its asymmetric positive and negative pulse fields, which renders discussed scheme
%available while a symmetric pulse field only gives rise to zero contribution (see the parameter $\gamma$).

%
%The work is supported by the cluster of excellence
%"Nanostructured Materials" of the state Saxony-Anhalt.

\section{Conclusions}
In summary we studied theoretically the photo-induced
spin dynamics in a quantum wire with a Rashba spin orbit
interaction. For an efficient and a  sub-picosecond control of the spin dynamics
the  pulses have to be shaped appropriately such that in effect a
linear momentum boost is transferred to the charge carrier.
For linear polarized photons we find that if the photon polarization axis
is  along  the wire's direction,
 the phase coherence in different spin channels is maintained, even though the charge carriers
 are speeded up.
In the case that the  photon pulse  polarization  is
perpendicular to the  wire we predict  a  spin
precession
comparable to that  induced by the Rashba SOI. The photon-induced
precession is   tunable in a considerable range
by scanning  the parameters of the pulse electric field.

\begin{acknowledgments}{
This work was supported by
the cluster of excellence "Nanostructured Materials" of the state
Saxony-Anhalt, and the DFG, Germany.}
\end{acknowledgments}

%\newpage

%\newpage\textbf{FIGURE CAPTIONS}\\
%Fig. 1 (Color online) ....


\begin{thebibliography}{99}

\bibitem{fukuda} M. Fukuda, \textit{Optical semiconductor
devices}, John Wiley \& Sons, Inc. 1999.

\bibitem{wolf} S. A. Wolf,
 D. D. Awschalom, R. A. Buhrman, J. M. Daughton, S.
von Moln\'{a}r, M. L. Roukes, A. Y. Chtchelkanova, and D. M. Treger,
Science \textbf{294}, 1488 (2001).

\bibitem{zutic} I. \v{Z}uti\'{c}, J. Fabian, and S. Das Sarma,
Rev. Mod. Phys. \textbf{76}, 323 (2004).

\bibitem{neamen} D. A. Neamen, \textit{Semiconductor physics and devices:
basic principles}, third edition, McGraw-Hill Companies, Inc., New
York, NY 2003.

\bibitem{spin} \emph{ Spin Electronics}, eds. D. D. Awschalom, %\emph{et al.},
 R. A. Buhrman, J. M. Daughton,  S. von Molnar,   M. L. Roukes
(Kluwer Academic Publishers, 2003);  D. D. Awschalom, N. Samarth,  Physics \textbf{2}, 50 (2009);
   %
    G. Malinowski,
     F. D. Longa, J. H. H. Rietjens, P. V. Paluskar, R. Huijink, H. J. M. Swagten and  B. Koopmans,
    Nature Physics \textbf{4}, 855 (2008);
    A. V. Kimel,
      A. Kirilyuk, P. A. Usachev, R. V. Pisarev, A. M. Balbashov, and Th. Rasing,
    Nature \textbf{435}, 655 (2005);
     G. P. Zhang, W. H\"{u}bner, G. Lefkidis, Y. Bai and  T. F. George,
      Nature Physics \textbf{5}, 499 (2009).
%
\bibitem{zhu08} Z.-G. Zhu, and J. Berakdar, Phys. Rev. B \textbf{77},
235438 (2008); J. Phys.:Condens. Matter \textbf{21}, 145801
(2009); Phys. Stat. Sol. B \textbf{247}, 641 (2010); A. Matos-Abiague, J. Berakdar
Phys. Rev. A \textbf{68},
063411   (2003).


%\bibitem{zhu09} Z.-G. Zhu, and J. Berakdar, J. Phys.:Condens. Matter
%\textbf{21}, 145801 (2009).

\bibitem{datta} S. Datta, and B. Das, Appl. Phys. Lett. \textbf{56}, 665 (1990).

\bibitem{koo} H. C. Koo, J. H. Kwon, J. Eom, J. Chang, S. H. Han, and M. Johnson, Science \textbf{325}, 1515 (2009).

\bibitem{tielking} N. E. Tielking, T. J. Bensky, and R. R. Jones, Phys. Rev. A
\textbf{51}, 3370 (1995).

\bibitem{hcp1} D. You, R. R. Jones, and P. H. Bucksbaum, and D. R. Dykaar, Opt. Lett. \textbf{18}, 290 (1993);
R. R. Jones, D. You, and P. H. Bucksbaum, Phys. Rev. Lett.
\textbf{70}, 1236 (1993); T. J. Bensky, G. Haeffler, and R. R. Jones, \textit{ibid}. \textbf{79}, 2018 (1997).

\bibitem{hcp2} A. E. Kaplan, Phys. Rev. Lett. \textbf{73}, 1243 (1994); A. E. Kaplan and P. L. Shkolnikov, \textit{ibid}. \textbf{75}, 2316 (1995).



\bibitem{matos}A. Matos-Abiague, and J. Berakdar,
Phys. Rev. Lett. \textbf{94}, 166801 (2005); Phys. Rev. B \textbf{70}, 195338  (2004);
 EPL \textbf{69}, 277 (2005).
% in \textit{Current
%Topics in Atomic, Molecular and Optical Physics}, edited by C. Sinha
%and S. Bhattacharyya (World Scienctific, London, 2007).
%
%
\bibitem{rashba}
E. I. Rashba, Fiz. Tverd. Tela (Leningrad) \textbf{2}, 1224 (1960)
[Sov. Phys. Solid State \textbf{2}, 1109 (1960)];Y. A. Bychkov and
E. I. Rashba, J. Phys. C \textbf{17}, 6039 (1984).
%
\bibitem{winkler} R. Winkler, \textit{Spin-Orbit coupling effects in two-dimensional electron and hole
systems}, (Springer, Berlin, 2003).

\bibitem{luo} J. Luo, H. Munekata, F. F. Fang, and P. J. Stiles, Phys. Rev. B
\textbf{41}, 7685 (1990).

\bibitem{das} B. Das, D. C. Miller, S. Datta, R. Reifenberger, W. P. Hong, P. K. Bhattacharya, J. Singh, and M. Jaffe, Phys. Rev. B \textbf{39}, 1411 (1989).


\bibitem{moskalenko} A. S. Moskalenko, A. Matos-Abiague, and J.
Berakdar, Phys. Rev. B \textbf{74}, 161303(R) (2006).

\bibitem{corkum} E. Dupont,
 P. B. Corkum, H. C. Liu, M. Buchanan, and Z. R. Wasilewski,
  Phys. Rev. Lett. \textbf{74}, 3596 (1995);
  H. M. van Driel, J. E. Sipe, and A. L. Smirl, Phys. Stat. Sol. B \textbf{243},  2278 (2006);
  C. Ruppert, S. Thunich, G. Abstreiter, A. Fontcuberta i Morral, A. W. Holleitner and M. Betz,
  Nano Lett., \textbf{10}, 1799 (2010).

  %
%\bibitem{winkler2000} R. Winkler, S. J. Papadakis, E. P. De Poortere, and M. Shayegan,
%Phys. Rev. Lett. \textbf{84}, 713 (2000).

%\bibitem{zhu} Zhen Gang Zhu, et al., Phys. Lett. A \textbf{300}, 658
%(2002); ibid. \textbf{306}, 249 (2003); Phys. Rev. B \textbf{68},
%224413 (2003).

%\bibitem{zhu1} Zhen-Gang Zhu, et al., Phys. Rev. B \textbf{70},
%174403 (2004).



%\bibitem{dion} C. M. Dion, A. Keller, and O. Atabek, Eur. Phys. J.
%D. \textbf{14}, 249 (2001).





\end{thebibliography}
\end{document}